\documentclass[         %
aps,                    
prd,                    
showpacs,               
nofootinbib,            
showkeys,               %
preprintnumbers,        %
floatfix]               
{revtex4}               
\usepackage{graphicx,longtable}

\begin{document}
\title{Neutrino-Neutrino Interactions and Flavor Mixing in Dense Matter}
\author{        A.~B. Balantekin}
\email{         baha@physics.wisc.edu}
\author{        Y. Pehlivan}
\email{         yamac@physics.wisc.edu}
\affiliation{  Department of Physics, University of Wisconsin\\
               Madison, Wisconsin 53706 USA }
\date{\today}
\begin{abstract}
An algebraic approach to the neutrino propagation in dense media is
presented. The Hamiltonian describing a gas of neutrinos interacting
with each other and with background fermions is written in terms of the
appropriate SU(N) operators, where N is the number of neutrino flavors.
The evolution of the resulting many-body problem is formulated as a
coherent-state path integral. Some commonly used approximations are
shown to represent the saddle-point solution of the path integral for
the full many-body system.
\end{abstract}
\medskip
\preprint{}
\maketitle

\vskip 1.3cm

\section{Introduction}

Neutrino propagation in dense matter \cite{Prakash:2001rx,Balantekin:1998yb}
is encountered near the core of a core-collapse supernovae
\cite{Qian:1994wh,Qian:1995ua,Pastor:2002we,Balantekin:2004ug,Fuller:2005ae,Strack:2005ux},
in the Early Universe \cite{Abazajian:2002qx,Ho:2005vj}, and possibly in
the gamma-ray bursts \cite{Kneller:2004jr}.
In particular, neutrino interactions play a crucial role in core-collapse
supernovae.
Neutrino oscillations in a core-collapse
supernova differ from the matter-enhanced neutrino oscillations in
the Sun as in the former there are additional effects coming from both
neutrino-neutrino scattering
\cite{Qian:1994wh,Pantaleone:1992eq,Pantaleone:1992xh} and
antineutrino flavor transformations \cite{Qian:1995ua}.
Exact integration of the neutrino evolution equations with such terms 
in the supernova environment turns out to be a very difficult problem.
A ``mean-field'' type approximation was proposed in Ref.
\cite{Qian:1994wh} which was adopted in exploratory
calculations of the conditions for r-process nucleosynthesis
in supernovae \cite{Pastor:2002we,Balantekin:2004ug}.
Analytical solutions were investigated in the limiting cases where
off-diagonal terms dominate \cite{Fuller:2005ae}. In addition, various
collective effects were explored
\cite{Kostelecky:1994dt,Samuel:1995ri,Friedland:2003dv,Friedland:2003eh,Duan:2005cp,Friedland:2006ke,Duan:2006an}.

The purpose of this article is to formulate the problem of neutrino
propagation in dense media algebraically. An algebraic formulation of the
problem should make hidden 
symmetries evident and also provide a framework to look for exact
solutions or systematic approximations. In the next section we introduce the
second-quantized formalism and the appropriate SU(2) algebra for two flavors.
In this section we also show that the evolution operator for the
standard MSW problem of neutrinos,
mixing with each other and interacting with background electrons, can be
written down exactly using an algebraic ansatz. In this and subsequent sections
we utilize a path-integral approach to the underlying many-body problem,
the details of which are sketched in the Appendix.
In Section III, we introduce the
neutrino-neutrino interaction and elucidate its algebraic nature. These
arguments are expanded into the situations where antineutrinos are
present in Section IV. The full problem with three flavors of both
neutrinos and antineutrinos is discussed in Section V.
Finally, a discussion of our results
in Section VI concludes the paper.

\section{Second-quantized formalism}

In our presentation we find it convenient to use the
second-quantized form \cite{Raffelt:1991ck}. For simplicity in
this section we assume that only
two flavors of neutrinos mix, which we take to be the
electron neutrino, $\nu_e$, and a combination of muon and tau
neutrinos, which we denote by $\nu_x$. (In the limit
$\theta_{13}=0$, one particular combination of the mu and tau
flavors decouple and the results in this section become exact with
$\nu_x$ being the combination orthogonal to the decoupled one
\cite{Balantekin:1999dx}). We first consider a situation where
there are no antineutrinos.
(We relax both of these assumptions in the subsequent sections).
The Hamiltonian describing mixing and
interaction of the neutrinos with the background electrons is
given by
\begin{eqnarray}\label{a1}
H_{\nu} &=& \frac{1}{2}\int d^3p \left( \frac{\delta
m^2}{2p}\cos{2\theta} - \sqrt{2} G_F N_e \right)
(a_x^\dagger(p)a_x(p)-a_e^\dagger(p)a_e(p)) \nonumber
\\ &+& \frac{1}{2}\int d^3 p \> \frac{\delta m^2}{2p} \> \sin{2\theta}
(a_x^\dagger(p)a_e(p)+a_e^\dagger(p)a_x(p)),
\end{eqnarray}
where $a_e^\dagger(p)$ and $a_x^\dagger(p)$ are the creation
operators for the left-handed
$\nu_e$ and $\nu_x$ with momentum p, respectively, and
$a_e(p)$ and $a_x(p)$ are the corresponding annihilation
operators. In Eq. (\ref{a1}), $\sqrt{2} G_F  N_e (x) = V_c$ is the
Wolfenstein potential describing the interaction of neutrinos with
electrons in neutral, unpolarized matter \cite{Wolfenstein:1977ue};
$N_e=n_{e^-}-n_{e^+}$ is the net electron density;
$\theta$ is the vacuum mixing angle; and
$\delta m^2 = m_2^2 - m_1^2$. In writing Eq. (\ref{a1}) we omitted a term
proportional to the identity (this includes the other Wolfenstein potential
$V_n (x) =  - (1/\sqrt{2}) G_F N_n(x)$
describing the neutral-current interaction of neutrinos with neutrons).
The presence of the Mikheev, Smirnov, Wolfenstein (MSW) resonance
\cite{Mikheev:1986gs,Mikheev:1986wj,Wolfenstein:1977ue} is manifest in
the first term.

The inherent SU(2) symmetry of the problem can be implemented by
the operators
\begin{equation}\label{a2}
J_+(p)= a_x^\dagger(p) a_e(p), \ \ \ \
J_-(p)=a_e^\dagger(p) a_x(p), \ \ \ \
J_0(p)=\frac{1}{2}\left(a_x^\dagger(p)a_x(p)-a_e^\dagger(p)a_e(p)
\right) , 
\end{equation}
which satisfy the commutation relations
\begin{equation}
\label{a3}
[J_+(p),J_-(q)] = 2 \delta^3(p-q)J_0(p), \ \ \
[J_0(p),J_\pm(q)] = \pm \delta^3(p-q)J_\pm(p).
\end{equation}
These equations describe as many commuting SU(2) algebras as the
number of distinct values of the neutrino momenta $p$ permitted by
the appropriate physical situation. Each $J(p)$ is realized in the
$j=1/2$ representation
due to the fermionic nature of neutrinos. The global
SU(2) operators, 
\begin{equation}
\label{a7} {\cal J}_{\pm} = \int d^3p J_{\pm}(p), \ \ \ \ {\cal
J}_0 = \int d^3p J_0(p) , 
\end{equation}
also satisfy the SU(2) commutation relations and play an
important role in the problem. Representation of the algebra
spanned by the operators ${\cal J}$ of Eq. (\ref{a7}) is obtained
by adding $N$ different copies of SU(2) each with $j=1/2$
where $N$ is the number of allowed values of neutrino momenta.

In terms of the SU(2) generators of Eq.(\ref{a2}), we can
write the neutrino Hamiltonian in Eq. (\ref{a1}) as follows:
\begin{equation}\label{a4}
H_{\nu} = \int d^3p \frac{\delta m^2}{2p} \left[ \cos{2\theta}
J_0(p) + \frac{1}{2} \sin{2\theta} \left(J_+(p)+J_-(p)\right)
\right] - \sqrt{2} G_F  \int d^3p N_e J_0(p) .
\end{equation}
Note that in the last term of Eq. (\ref{a4}) we kept the electron
density inside the momentum integral since it depends on
the direction of the neutrino momentum (neutrino traveling in
different directions will, in principle, see different electron
densities), but of course not on the
absolute value.

In general we are interested in finding the operator $U$, describing the
evolution of the system:
\begin{equation}
\label{a5}
i\frac{\partial U}{\partial t} = H_{\nu} U.
\end{equation}
In most cases, however, the physical interest is in finding the
evolution of a particular state of the system. One may start, for
example, with a state in which all permitted electron neutrino
states are occupied and all $\nu_x$ states
are empty i.e.,
\begin{equation}
\label{a6} |\phi \rangle =\prod_{p\in {\mathcal P}}
a_e^\dagger(p)|0 \rangle , 
\end{equation}
where $|0 \rangle $ is the particle vacuum. 
${\mathcal P}$ denotes the set of all allowed
neutrino momenta. The state $|\phi
\rangle $ is annihilated by the operators $J_-(p)$ for all $p$ and
also by ${\cal J}_-$. It is the lowest-weight eigenstate of the
representation with $j=N/2$ of the global SU(2) operators ${\cal
J}_0$ and ${\cal J}_\pm$.

The evolution operator can be found by employing the unitary
ansatz
\begin{equation}
\label{a8} U = \exp \left( \int d^3p \tau(p,t) J_+(p) \right) \exp
\left( \int d^3p \log (1+|\tau(p,t)|^2) J_0(p) \right)\exp \left(
-\int d^3p \tau^*(p,t) J_-(p) \right).
\end{equation}
Here $\tau(p,t)$ is a function to be determined by substituting
this ansatz into Eq. (\ref{a5}). One can differentiate $U$ of
Eq. (\ref{a8})  using the operator chain rule. Differentiation 
introduces the operator 
$J_0(p')$ between the first and second exponentials, and the operator 
$J_-(p')$ between the second and third exponentials. Since, for example, 
$J_0(p')$ does not commute with the first exponential, to write this operator 
before the first exponential, one needs to introduce the identity 
operator in the form $\exp \left( \int d^3p \tau(p,t) J_+(p) \right)
\exp \left( - \int d^3p \tau(p,t) J_+(p) \right)$ after the operator 
$J_0(p')$ and use the identity 
\[
\exp ({\cal O}_1) {\cal O}_2 \exp (-{\cal O}_1) = {\cal O}_2 + 
[{\cal O}_1, {\cal O}_2 ] + \frac{1}{2!} [ {\cal O}_1, [{\cal O}_1, 
{\cal O}_2 ]] + \cdots, 
\]
valid for any two arbitrary operators ${\cal O}_1$ and ${\cal O}_1$.  
Moving all such terms to the left of the exponentials we find 
\begin{eqnarray}
i\frac{\partial U}{\partial t} &=& \left[ \int d^3p \left(
i\frac{\dot{\tau}(p,t)\tau^*(p,t) -
\tau(p,t)\dot{\tau}^*(p,t)}{1+|\tau(p,t)|^2}
J_0(p)\right) \right. \nonumber \\
&+& \left. \int d^3p
\left(i\frac{\dot{\tau}(p,t)}{1+|\tau(p,t)|^2}J_+(p)
-i\frac{\dot{\tau}^*(p,t)}{1+|\tau(p,t)|^2}J_-(p)\right) \right] U,
\label{a9}
\end{eqnarray}
where the dot denotes derivative with respect to time. Eq.
(\ref{a5}) is satisfied if $\tau(p,t)$ obeys the equations:
\begin{equation}
i\frac{\dot{\tau}(p,t)\tau^*(p,t) -
\tau(p,t)\dot{\tau}^*(p,t)}{1+|\tau(p,t)|^2}
=\frac{\delta m^2}{2p}\cos{2\theta} - \sqrt{2} G_F N_e,
\label{a10}
\end{equation}
and
\begin{equation}
\label{a11}
i\frac{\dot{\tau}(p,t)}{1+|\tau(p,t)|^2}
=-i\frac{\dot{\tau}^*(p,t)}{1+|\tau(p,t)|^2}=\frac{1}{2}
\frac{\delta m^2}{2p}\sin{2\theta} ,
\end{equation}
i.e. equations for different neutrino momenta decouple as expected.
In order to satisfy the condition $U(t=0)=1$, $\tau(p,t=0)$ should be
zero for all $p$.
Using Eqs. (\ref{a10}) and (\ref{a11}), one can show that
$\tau(p,t)$ is the solution of the following nonlinear (Riccati-type)
first-order differential equation:
\begin{equation}
\label{a12}
i\dot{\tau}(p,t)=\frac{1}{2}\frac{\delta
m^2}{2p}\sin{2\theta}(1- \tau(p,t)^2) + \left(\frac{\delta
m^2}{2p}\cos{2\theta} - \sqrt{2} G_F N_e\right) \tau(p,t).
\end{equation}
$\tau(p,t)$ can be interpreted as the ratio of the one-body
neutrino wave functions
\begin{equation}
\label{a13} \tau(p,t)=\frac{\psi_x(p,t)}{\psi_e(p,t)}.
\end{equation}
To see this, one only needs to substitute Eq. (\ref{a13}) into Eq.
(\ref{a12}) together with the normalization condition
\begin{equation}
|\psi_e|^2+|\psi_x|^2=1.
\end{equation}
This way, it becomes clear that Eq. (\ref{a12}) is equivalent to
the Schrodinger equation
\begin{equation}
\label{a14} i\frac{\partial}{\partial t} \left( \begin{array}{c}
\psi_e
\\ \psi_x \\ \end{array} \right) =
\frac{1}{2}\left(%
\begin{array}{cc}
  A-\Delta\cos{2\theta} &\Delta\sin{2\theta} \\
  \Delta\sin{2\theta} & -A+\Delta\cos{2\theta} \\
\end{array} \right)
\left( \begin{array}{c} \psi_e \\
\psi_x \\ \end{array} \right)
\end{equation}
where we defined
\begin{equation}
\label{a15} \Delta=\frac{\delta m^2}{2p}, \ \ \ \ \ \ \ A=\sqrt{2}
G_F N_e .
\end{equation}
For $\tau (t=0)$ to be zero, we need the initial conditions
$\psi_x(t=0)=0$ and $\psi_e (t=0) =1$. Eq. (\ref{a14}) is nothing
but the standard MSW evolution equation for two flavors. Note that
our ansatz for the evolution operator provides the ``coset-space''
formulation of this problem (cf. Eq. (\ref{a12}) with the
formulation in the Appendix of Ref. \cite{Balantekin:2004tk}). The
initial state $| \phi \rangle$, given in Eq. (\ref{a6}), then
evolves into the state
\begin{equation}
\label{a16} \exp \left( - \frac{1}{2} \int_{{\mathcal P}} d^3p
\log (1+|\tau(p,t)|^2) \right) \exp \left( \int_{{\mathcal P}}
d^3p \tau(p,t) J_+(p) \right) |\phi \rangle ,
\end{equation}
where $\tau (p,t)$ is obtained by solving Eq. (\ref{a12}). 
Both the initial state in
Eq. (\ref{a6}) and evolved state in Eq. (\ref{a16}) are normalized
to unity. The evolved state in Eq. (\ref{a16}) is the standard
normalized SU(2) coherent state.

\section{Neutrino-Neutrino Interactions}

In some astrophysical environments, such as supernovae and the 
Early Universe, where neutrino density can become very large
\cite{Balantekin:2003ip},
the Hamiltonian in Eq. (\ref{a1}) is no longer sufficient.
Neutrino self interactions must also be taken into account.
In most cases one is mainly interested in the forward scattering of
neutrinos from other neutrinos. These are described by the
Hamiltonian
\begin{eqnarray}\label{b1}
H_{\nu \nu} &=& \frac{G_F}{\sqrt{2}V}  \int d^3p \>d^3q \> R_{pq}
\> \left[ a_e^\dagger(p)a_e(p)a_e^\dagger(q)a_e(q)
 +a_x^\dagger(p)a_x(p)a_x^\dagger(q)a_x(q) \right. \nonumber \\
&+& \left. a_x^\dagger(p)a_e(p)a_e^\dagger(q)a_x(q) +
a_e^\dagger(p)a_x(p)a_x^\dagger(q)a_e(q) \right]
\end{eqnarray}
where we defined
\begin{equation}
R_{pq} = (1-\cos\vartheta_{pq}) . \label{b2}
\end{equation}
Here, $\vartheta_{pq}$ is the angle between the momentum
directions of the neutrinos with momenta ${\bf p}$ and ${\bf q}$ and
we used box quantization conditions for a box with volume $V$.
In forward scattering, neutrinos exchange their momenta by
exchanging a $Z$ boson with momentum ${\bf p}-{\bf q}$ or a $Z$
boson with zero momentum. The coefficient $1-\cos\vartheta_{pq}$
ensures that the neutrinos which move parallel to each other
do not undergo forward scattering.

Numerical studies indicate that the addition of a neutrino
background results in very interesting physical effects such as
coherent flavor transformation. For details see, for example, Refs.
\cite{Qian:1994wh,Qian:1995ua,Pastor:2002we,Balantekin:2004ug,Fuller:2005ae}
and
\cite{Friedland:2003dv,Friedland:2003eh,Duan:2005cp,Friedland:2006ke,Duan:2006an}. 
The major difficulty in studying the effects
of a neutrino background on flavor evolution is the inherent
nonlinearity of the problem.
On the other hand, the contribution from $H_{\nu \nu}$ is
relatively easy to study in the algebraic formalism. First, one
needs to write $H_{\nu \nu}$ given in Eq. (\ref{b1}) in terms of
the SU(2) generators of Eq. (\ref{a2}). Omitting a term
proportional to the identity, we find that
\begin{equation}\label{b3}
H_{\nu \nu} = \frac{\sqrt{2} G_F}{V} \int d^3p \> d^3q \>  R_{pq}
\> {\bf J}(p) \cdot {\bf J}(q) .
\end{equation}
Here, ${\bf J}(p) \cdot {\bf J}(q)$ is defined as
\begin{equation} \label{b4}
{\bf J}(p) \cdot {\bf J}(q) = J_0(p)J_0(q) + \frac{1}{2}
J_+(p)J_-(q) + \frac{1}{2} J_-(p)J_+(q) .
\end{equation}
This expression is very similar, but not identical, to the global
SU(2) Casimir operator. In fact, $H_{\nu \nu}$ commutes with the global
SU(2) operators of Eq. (\ref{a7})
\begin{equation} \label{b6}
[ H_{\nu \nu}, {\cal J}_i] = 0
\end{equation}
although the Hamiltonian $H_{\nu}$ does not:
\begin{equation}
[ H_{\nu}, {\cal J}_i] \neq 0 .
\label{b5}
\end{equation}
On the other hand, the last term of $H_{\nu}$ in Eq. (\ref{a4}) is
proportional to ${\cal J}_0$ for a constant electron density.
Thus if $\delta m^2$ were zero the
evolution problem for $H_{\nu} + H_{\nu \nu}$ could have been
solved in the ${\cal J}$-basis for a constant electron density.
We see that, although $H_{\nu
\nu}$ presents a difficulty in numerical studies, it is relatively
easier to work with in this algebraic approach. The
noncommutativity of the first integral in Eq. (\ref{a4}) with
generators of the total angular momentum operators ${\cal J}$,
however, is the salient difficulty of the algebraic approach to
the problem.

One possible approach to the problem of finding the evolution
operator with $H_{\nu} + H_{\nu \nu}$ is to seek a path integral
representation. One can appropriately use the SU(2) coherent
states
\begin{equation}
\label{b7} |z(t)\rangle = N \exp{\left(\int_{{\mathcal P}} d^3p 
\> z(p,t) J_+(p) \right)} |\phi \rangle.
\end{equation}
Here $|\phi\rangle$ is the state defined in Eq. (\ref{a6}) and the
normalization constant $N$ is given by
\begin{equation}
N=\exp \left( - \frac{1}{2} \int_{{\mathcal P}} d^3p \log
(1+|z(p,t)|^2) \right) .
\end{equation}
Clearly when $z(p,t)=\tau (p,t)$ the state in Eq. (\ref{b7})
becomes the exact solution of the evolution with $H_{\nu}$ alone.
Path integral representation of the matrix element of the
evolution operator calculated with $H_{\nu} + H_{\nu \nu}$ between
two states $|z(t_i) \rangle $ and $|z'(t_f) \rangle$ is given by
\begin{equation}
\label{b7a} \langle z'(t_f)|U|z(t_i) \rangle = \int D[z,z^*] \,
e^{iS[z,z^*]}
\end{equation}
where the path integral measure is
\begin{equation}
\label{b7b}
D[z,z^*]=\lim_{N\to\infty}%
e^{-2\sum_{\alpha=1}^N\int_{{\mathcal P}} dp %
\log \left(1+|z(p,t_\alpha)|^2\right)} \prod_{\alpha=1}^N
\prod_{p\in{\mathcal P}}2!\frac{d^2z(p,t_\alpha)}{2\pi i} . 
\end{equation}
Here, the exponential factor arises because the $SU(2)$ coherent
states are overcomplete and therefore require a weight function in
the resolution of identity. A detailed derivation can be found in
the Appendix. In the above formula $dz^2$ refers to $dz dz^*$.
The action functional is given by\footnote{We use the
shorthand notation $\langle {\mathcal O} \rangle
=\langle z(t)|{\mathcal O}|z(t) \rangle $}
\begin{equation}
\label{b8} S[z,z^*]= \int_{t_i}^{t_f}dt \langle i\frac{\partial}{
\partial t}-H_\nu-H_{\nu\nu} \rangle -i \log
\langle z'(t_f)|z(t_f) \rangle.
\end{equation}

The leading contribution to this path integral comes from the
stationary path $|z(t) \rangle$ which minimizes the action
functional. As in any variational problem, stationary path can be
found by solving the Euler-Lagrange equations
\begin{equation}
\label{b9} \left(\frac{d}{dt}\frac{\partial}{\partial \dot{z}}-
\frac{\partial}{\partial z}\right)L(z,z^*)=0 , \ \ \ \ \ \
\left(\frac{d}{dt}\frac{\partial}{\partial \dot{z}^*}-
\frac{\partial}{\partial z^*}\right)L(z,z^*)=0 ,
\end{equation}
where
\begin{equation}
\label{b10} L[z,z^*]=\langle i\frac{\partial}{\partial
t}-H_\nu-H_{\nu\nu} \rangle
\end{equation}
which plays the role of the Lagrangian. To solve the
Euler-Lagrange equations, however, we first need to calculate the
Lagrangian as a function of $z$ and $z^*$. This can be done by
substituting $H_\nu$ from Eq. (\ref{a4}) and $H_{\nu\nu}$ from Eq.
(\ref{b3}) into the Lagrangian of Eq. (\ref{b10}) and using the
expectation values 
\begin{equation}
\label{b11} \langle J_+(p) \rangle = \langle J_-(p) \rangle ^*=
\frac{z^*(p,t)}{1+|z(p,t)|^2} , \quad\quad \langle J_0(p) \rangle
= -\frac{1}{2}\ \frac{1-|z(p,t)|^2}{1+|z(p,t)|^2} ,
\end{equation}
which are valid for $p\in{\mathcal P}$. When $p\not\in{\mathcal P}$, 
however, these 
expectation values are equal to zero. When $p\neq q$
coherent states also obey
\begin{equation}\label{b14}
\langle J_a(p)J^\dagger_a(q) \rangle =\langle J_a(p)
\rangle\langle J^\dagger_a(q) \rangle
\end{equation}
for $a=0,\pm$. Using above formulas and Eqs. (\ref{a4}),
(\ref{b10}) and (\ref{b3}) together with
\begin{equation}
\langle i\frac{d}{dt}\rangle =i \int
dp\frac{\dot{z}(p)z^*(p)}{1+|z(p)|^2} ,
\end{equation}
we obtain the following expression for the Lagrangian:
\begin{eqnarray} \label{b15}
L[z,z^*] &=& L_\nu - %
\frac{1}{4}\sqrt{2}G_F \int_{{\mathcal P}} d^3p
\frac{B(p)(1-|z(p)|^2)+2B_{ex}(p)z(p)+2B_{xe}(p)z^*(p)}{1+|z(p)|^2}
.
\end{eqnarray}
Here we defined
\begin{eqnarray}\label{b15a}
L_\nu &=& i\int_{{\mathcal P}} d^3p \frac{\dot{z}(p,t)
z^*(p,t)}{1+|z(p,t)|^2} + \frac{1}{2} \int_{{\mathcal P}} d^3p
\left( \frac{\delta m^2}{2p}\cos{2\theta} - \sqrt{2} G_F
N_e \right) \frac{1-|z(p,t)|^2}{1+|z(p,t)|^2} \nonumber\\
&-&\frac{1}{2} \int_{{\mathcal P}} d^3p \frac{\delta
m^2}{2p}\sin{2\theta} \frac{z(p,t)+z^*(p,t)}{1+|z(p,t)|^2}
\end{eqnarray}
and
\begin{equation}\label{b15b}
B(p)=\frac{\sqrt{2}G_F}{V} \int_{{\mathcal P}} d^3q
R_{pq}\frac{1-|z(q,t)|^2}{1+|z(q,t)|^2} , \quad\quad\quad
B_{xe}(p)=B^*_{ex}=2\frac{\sqrt{2}G_F}{V} \int_{{\mathcal P}}
d^3q R_{pq}\frac{z(q,t)}{1+|z(q,t)|^2} .
\end{equation}
Euler-Lagrange equations which follow from this Lagrangian are
\begin{equation}\label{b16}
i\dot{z}(p,t)=\beta(p,t)-\alpha(p,t) z(p,t)-\beta^*(p,t) z(p,t)^2
\end{equation}
and its complex conjugate. The coefficients $\alpha$ and $\beta$
are given by:
\begin{equation}
\label{b17} \alpha(p,t) = -\frac{\delta m^2}{2p}\cos{2\theta}+
\sqrt{2}G_F N_e + B(p) , \quad\quad \beta(p,t)=\frac{1}{2}
\frac{\delta m^2}{2p}\sin{2\theta} + \frac{1}{2}B_{xe} (p) .
\end{equation}
As in the previous section, $z(p,t)$ can be interpreted as the
ratio of the one-body neutrino wave functions $\psi_e$, $\psi_x$:
\begin{equation}
\label{b19}
z(p,t)=\frac{\psi_x(p,t)}{\psi_e(p,t)}
\end{equation}
with the normalization condition
\begin{equation}
\label{b20} |\psi_e|^2+|\psi_x|^2=1 .
\end{equation}
If we substitute Eq. (\ref{b19}) into Eq.(\ref{b16}), we see that
Eq. (\ref{b16}) is equivalent to the Schrodinger equation
\begin{equation}
\label{b21} i\frac{\partial}{\partial t} \left( \begin{array}{c}
\psi_e
\\ \psi_x \\ \end{array} \right) =
\frac{1}{2}\left(%
\begin{array}{cc}
  A+B-\Delta\cos{2\theta} & B_{ex}+\Delta\sin{2\theta} \\
  B_{xe}+\Delta\sin{2\theta} & -A-B+\Delta\cos{2\theta} \\
\end{array} \right)
\left( \begin{array}{c} \psi_e \\
\psi_x \\ \end{array} \right)
\end{equation}
where $\Delta$ and $A$ are defined in Eq. (\ref{a15}). $B$ and
$B_{ex}$ are obtained by substituting Eq. (\ref{b19}) into Eq.
(\ref{b15b}):
\begin{equation}
\label{b23} B=\frac{\sqrt{2}G_F}{V} \int_{{\mathcal P}} d^3q
R_{pq}\left[|\psi_e(q,t)|^2-|\psi_x(q,t)|^2\right],
\end{equation}
\begin{equation}
\label{b24} B_{ex}=\frac{2\sqrt{2}G_F}{V} \int_{{\mathcal P}}
d^3q R_{pq}\psi_e(q,t)\psi^*_x(q,t).
\end{equation}
Eq. (\ref{b21}) is commonly used to study neutrino propagation with
neutrino-neutrino interactions (see Refs. \cite{Qian:1994wh}
through \cite{Duan:2005cp}). Here we illustrated that it is not an
exact result, but represents the saddle point solution to the
many-body problem.

There is an alternative method to obtain the approximate results
outlined above. This method, typically employed in the random-phase
approximation approach to many-body problems, approximates 
the product of two
commuting arbitrary operators $\hat{\cal O}_1$ and $\hat{\cal O}_2$ as
\begin{equation}
\label{rpa}
\hat{\cal O}_1 \hat{\cal O}_2 \sim \hat{\cal O}_1 \langle \xi |
\hat{\cal O}_2 | \xi \rangle + \langle \xi | \hat{\cal O}_1 |
\xi \rangle \hat{\cal O}_2 -
\langle \xi | \hat{\cal O}_1 |
\xi \rangle
\langle \xi | \hat{\cal O}_2 |
\xi \rangle ,
\end{equation}
provided that the condition
\begin{equation}
\label{rpa1}
\langle \xi | \hat{\cal O}_1  \hat{\cal O}_2 | \xi \rangle =
\langle \xi | \hat{\cal O}_1 |
\xi \rangle
\langle \xi | \hat{\cal O}_2 |
\xi \rangle
\end{equation}
is satisfied. In Eq. (\ref{rpa}) the state $| \xi \rangle$  must
be appropriately chosen so that Eq. (\ref{rpa1}) is satisfied.
Since the quantity $R_{pq}$ is zero when the
operators do not commute (i.e., when ${\bf p=q}$),
one can apply this approximation technique to the quadratic element
${\bf J}(p)\cdot {\bf J}(q)$ which appear in the neutrino-neutrino
forward scattering Hamiltonian $H_{\nu\nu}$ given in Eq.
(\ref{b3}). It follows from Eq. (\ref{b14})
that the SU(2) coherent states
satisfy the condition stated in Eq. (\ref{rpa1}).
Using the symmetry
$R_{pq}=R_{qp}$, one can then replace $H_{\nu \nu}$ by
\begin{equation}
\label{apprpa}
{\mathcal H}_{\nu \nu} \sim 2\frac{\sqrt{2}G_F}{V} \int d^3p \> d^3q
\> R_{pq} \> \left( J_0(p) \langle J_0(q) \rangle + \frac{1}{2}
J_+(p) \langle J_-(q) \rangle + \frac{1}{2} J_-(p) \langle J_+(q)
\rangle\right) ,
\end{equation}
where the expectation values are calculated using the SU(2)
coherent states (the variable $z(t)$ of these coherent states
must be obtained in a self-consistent way). In writing Eq. 
(\ref{apprpa}) we omitted a term proportional to identity. 
The total Hamiltonian for the two neutrino flavors
propagating in the presence of electrons and other neutrinos in
the background becomes approximately
\begin{eqnarray}
H_{\nu}+{\mathcal H}_{\nu \nu} &\sim& \int d^3p \left(\frac{\delta
m^2}{2p}\cos{2\theta}- \sqrt{2} G_F N_e-B(p)\right) J_0(p)
 \\
&+&  \int d^3p \left(\frac{1}{2}
\sin{2\theta}+\frac{1}{2}B_{xe}(p)\right) J_+(p) + \int
d^3p\left(\frac{1}{2}
\sin{2\theta}+\frac{1}{2}B_{ex}(p)\right)J_-(p) . \nonumber
\end{eqnarray}
Here $B(p)$, $B_{ex}(p)$ and $B_{xe}(p)$ are given in Eq.
(\ref{b15b}) and we used the expectation values given in Eq.
(\ref{b11}). Since the above approximate
Hamiltonian is linear in the SU(2)
generators we can use the same ansatz for the evolution operator
of the system as we did in section II, \emph{i.e.}
\begin{equation}
U = \exp \left( \int d^3p z(p,t) J_+(p) \right) \exp \left( \int
d^3p \log (1+|z(p,t)|^2) J_0(p) \right)\exp \left( -\int d^3p
z^*(p,t) J_-(p) \right).
\end{equation}
It is straightforward to show that, substituting this ansatz into
the equation
\begin{equation}
i\frac{d}{dt}U=\left(H_{\nu}+{\mathcal H}_{\nu \nu}\right)U
\end{equation}
yields nothing but Eq. (\ref{b16}) for $z(p,t)$. It is also
straightforward to generalize this linearization scheme to the
situations where antineutrinos and more flavors are present.

Corrections to the path integral will naturally arise as a result of 
the deviations from the classical (i.e., stationary) 
path given by Eq. (\ref{b16}) or Eq.
(\ref{b21}). To calculate the effect of small deviations one can
carry out a Taylor expansion of the action around the classical
path. It is helpful to introduce the following
nonlinear transformation in order to eliminate the exponential
factor in the path integral measure in Eq. (\ref{b7b}):
\begin{equation}\label{b26}
\beta(p,t)=\frac{z(p,t)}{\sqrt{1+|z(p,t)|^2}} \ \ \ \ \
\beta^*(p,t)=\frac{z^*(p,t)}{\sqrt{1+|z(p,t)|^2}}.
\end{equation}
The Jacobian resulting from this change of variables conveniently
cancels out the exponential factor. It is straightforward to show
that 
\begin{equation}\label{b27}
e^{-2\sum_{\alpha=1}^N\int_{{\mathcal P}} dp %
\log \left(1+|z(p,t_\alpha)|^2\right)} \prod_{\alpha=1}^N
\prod_{p\in{\mathcal P}}2!\frac{d^2z(p,t_\alpha)}{2\pi i}
=\prod_{\alpha=1}^N \prod_{p\in{\mathcal
P}}2!\frac{d^2\beta(p,t_\alpha)}{2\pi i}
\end{equation}
The path integral then takes the form
\begin{equation}\label{b28}
\langle z'(t_f)|U|z(t_i) \rangle = \int\lim_{N\to\infty}
\prod_{\alpha=1}^N \prod_{p\in{\mathcal
P}}\frac{d\beta(p,t_\alpha)d\beta^*(p,t_\alpha)}{i\pi}e^{iS[\beta,\beta^*]}
\end{equation}
where the action functional, written in terms of the new variables
is as follows:
\begin{eqnarray}\label{b29}
S[\beta,\beta^*]&=&\int_{t_i}^{t_f} dt \left[
\int dp \left( %
\frac{\beta(p,t)\dot{\beta}^*(p,t)-\dot{\beta}(p,t)\beta^*(p,t)}{2i}
+\frac{1}{2}\left(\frac{\delta m^2}{2p}\cos{2\theta}- \sqrt{2} G_F
N_e\right)\left(1-2|\beta(p,t)|^2\right) \right. \right. \nonumber \\
&-& \left. \frac{1}{2}\frac{\delta m^2}{2p}\sin{2\theta}
\left(\beta(p,t)+\beta^*(p,t)\sqrt{1-|\beta(p,t)|^2}\right)
\right) \\ %
&+& \frac{1}{2}\frac{\sqrt{2}G_F}{V} \int d^3p
\> d^3q \> R_{pq} \left[\left(|\beta(p,t)|^2+|\beta(q,t)|^2
-2|\beta(p,t)|^2|\beta(q,t)|^2\right) \right.\nonumber \\ &+&
\left. \left.
\sqrt{\left(1-|\beta(p,t)|^2\right)\left(1-|\beta(q,t)|^2\right)}
\left(\beta(p,t)\beta^*(q,t)+\beta^*(p,t)\beta(q,t)\right)
\right] \right] \nonumber
\end{eqnarray}
One obtains the following equation of motion (and its complex conjugate) 
from the variation
of this action:
\begin{eqnarray}\label{b30}
i\dot{\beta}(p,t)&=&\left( \left(\frac{\delta
m^2}{2p}\cos{2\theta}- \sqrt{2} G_F
N_e\right)-\frac{\sqrt{2}G_F}{V} \int \> d^3q \> R_{pq}
\left(1-2|\beta(q,t)|^2\right)\right)\beta (p,t)\\
&+&\left(\frac{\delta m^2}{2p}\sin{2\theta}+\frac{\sqrt{2}G_F}{V}
\int \> d^3q \> R_{pq}
\sqrt{1-|\beta(q,t)|^2}\beta (q,t)\right) 
\frac{3|\beta(p,t)|^2-2}{2\sqrt{1-|\beta(p,t)|^2}}\nonumber
\\
&+& \left(\frac{\delta m^2}{2p}\sin{2\theta}+\frac{\sqrt{2}G_F}{V}
\int \> d^3q \> R_{pq}
\sqrt{1-|\beta(q,t)|^2}\beta(q,t)\right) 
\frac{{\beta(p,t)}^2}{\sqrt{1-|\beta(p,t)|^2}}\nonumber
\end{eqnarray}
This classical path is the same as the one given in Eq.
(\ref{b16}). This can be shown directly by substituting the
transformation in Eq. (\ref{b26}) into Eq. (\ref{b16}). We will
denote the classical path as $\beta_{cl}$. Since the first order
variations are zero on the classical path by definition, the
Taylor expansion around $\beta_{cl}$ yields
\begin{eqnarray}\label{b31}
S[\beta,\beta^*]&=&S[\beta_{cl},\beta_{cl}^*]%
+\frac{1}{2}\left(\beta-\beta_{cl}\right)^T
\left(\frac{\delta^2 S} {\delta \beta \> \delta \beta}\right)_{cl}%
\left(\beta-\beta_{cl}\right) \\
&+& \left(\beta-\beta_{cl}\right)^T \left(\frac{\delta^2 S}{\delta
\beta \> \delta
\beta^*}\right)_{cl}\left(\beta^*-\beta^*_{cl}\right)%
+\frac{1}{2}\left(\beta^*-\beta^*_{cl}\right)^T\left(\frac{\delta^2
S}{\delta \beta^* \delta
\beta^*}\right)_{cl}\left(\beta^*-\beta^*_{cl}\right)+\dots
\nonumber
\end{eqnarray}
Here $(\dots )_{cl}$ indicates that the derivatives are to be
calculated on the classical path $\beta_{cl}$ and
$\left(\beta-\beta_{cl}\right)^T
\left(\frac{\delta^2 S} {\delta \beta \> \delta \beta}\right)_{cl}%
\left(\beta-\beta_{cl}\right)$ is a short hand notation for the
matrix product
\begin{equation}
\sum_{p,k}\sum_{q,m}\left(\beta(p,t_k)-\beta_{cl}(p,t_k)\right)^T
\left(\frac{\delta^2 S} {\delta \beta(p,t_k) \> \delta
\beta(p,t_m)}\right)_{cl}%
\left(\beta(p,t_m)-\beta_{cl}(p,t_m)\right),
\end{equation}
and similarly for the other terms. The sums over $p$ and $q$ run
through the allowed momentum modes and the sums over $k$ and $m$
run from $1$ to $N$ which is the number of time intervals we
introduced in path integral. $N\to\infty$ limit should be taken as
explained in the Appendix. For small deviations from the classical
path, one can ignore the higher order terms in the expansion
(\ref{b31}) and substitute it in Eq. (\ref{b28}):
\begin{equation}\label{b32}
\langle z'(t_f)|U|z(t_i) \rangle = e^{iS[\beta_{cl},\beta_{cl}^*]}
\int\lim_{N\to\infty} \prod_{k=1}^N \prod_{p\in{\mathcal
P}}\frac{d\tilde{\beta}(p,t_k)d\tilde{\beta}^*(p,t_k)}{i\pi}e^{%
i\left(\frac{1}{2}\tilde{\beta}^T\left(\frac{\delta^2 S}{\delta \beta \> \delta
\beta}\right)_{cl} \tilde{\beta}%
+\tilde{\beta}^T\left(\frac{\delta^2 S}{\delta \beta \> \delta
\beta^*}\right)_{cl}\tilde{\beta}^*%
+\frac{1}{2}{\tilde{\beta}^{*T}}\left(\frac{\delta^2 S}{\delta
\beta^* \delta \beta^*}\right)_{cl}{\tilde{\beta}^*}\right)}
\end{equation}
where we defined $\tilde{\beta}=\beta-\beta_{cl}$. The classical
action $S[\beta_{cl},\beta_{cl}^*]$ is taken out of the
integration since it does not depend on $\tilde{\beta}$. The
lowest order quantum corrections are captured by the Gaussian
integral in Eq. (\ref{b32}). The result of the integration is
\begin{equation}\label{b33}
\langle z'(t_f)|U|z(t_i) \rangle = \lim_{N\to\infty} (i\pi)^{N+P}
\frac{e^{iS[\beta_{cl},\beta_{cl}^*]}}{\sqrt{Det\left(KM-L^TK^{-1}L\right)}}.
\end{equation}
Here $P$ denote the number of allowed momentum modes. The matrices
$K,M$ and $L$ are given as follows:
\begin{equation}
K(p,k,q,m)=\frac{1}{2}\left(\frac{\delta^2 S}{\delta x(p,t_k) \>
\delta
x(q,t_m)}\right)_{cl}\ \ \ \ \ \  %
M(p,k,q,m)=\frac{1}{2}\left(\frac{\delta^2 S}{\delta y(p,t_k) \>
\delta y(q,t_m)}\right)_{cl}%
\end{equation}
\begin{equation}
L(p,k,q,m)=\frac{1}{2}\left(\frac{\delta^2 S}{\delta x(p,t_k) \>
\delta y(q,t_m)}\right)_{cl}
\end{equation}
where $x=(\tilde{\beta}+\tilde{\beta}^*)/2$ and
$y=(\tilde{\beta}-\tilde{\beta}^*)/2i$. The fundamental difficulty
involved in the calculation of the determinant in Eq. (\ref{b33})
is the existence of non-diagonal terms in the matrices $K,M$ and
$L$. These terms are generated by the $\int d^3 p \> d^3 q \>
R_{pq}\>\dots$ integral in the action given in Eq. (\ref{b29}).
A complete analysis of these determinants is beyond the scope of the 
present paper and will be given elsewhere. 

\section{Neutrino-Antineutrino Interactions}

Environments such as core-collapse supernovae and the Early Universe
contain copious amounts of antineutrinos as well as neutrinos.
When antineutrinos are added to the picture the effective flavor
evolution Hamiltonian becomes
\begin{equation}\label{c1}
H=H_\nu+H_{\bar{\nu}}+H_{\nu\nu}+H_{\bar{\nu}\bar{\nu}}+H_{\nu\bar{\nu}}
.
\end{equation}
Here, $H_\nu$ and $H_{\nu\nu}$ are given in Eqs. (\ref{a1}) and
(\ref{b1}). $H_{\bar{\nu}}$ and $H_{\bar{\nu}\bar{\nu}}$ are the
same as $H_\nu$ and $H_{\nu\nu}$, respectively except that the
neutrino operators $a$ and $a^\dagger$ are replaced by the
antineutrino operators $b$ and $b^\dagger$ and the sign of the
$N_e$ term is reversed. The neutrino-antineutrino forward
scattering Hamiltonian $H_{\nu\bar{\nu}}$ is given as follows:
\begin{eqnarray}\label{c2}
H_{\nu\bar{\nu}}=&-&\frac{\sqrt{2}G_F}{V}\int d^3pd^3q
R_{pq} \left[a_e^\dagger(p)a_e(p)b_e^\dagger(q)b_e(q)
+ a_x^\dagger(p)a_x(p)b_x^\dagger(q)b_x(q)\right.
\\
&+& \left. a_x^\dagger(p)a_e(p)b_x^\dagger(q)b_e(q) +
a_e^\dagger(p)a_x(p)b_e^\dagger(q)b_x(q)\right] . \nonumber
\end{eqnarray}

In addition to the neutrino operators defined in Eq. (\ref{a2}),
we now define the antineutrino SU(2) operators\footnote{Time-reversal
invariance requires that the order of the flavors is flipped
in the definition of the antineutrino algebra as compared with the
definition of the neutrino algebra in Eq. (\ref{a2}).}
\begin{equation}\label{c3}
\bar{J}_+(p)=b_{e}^\dagger(p)b_x(p) , \ \ \ \
\bar{J}_-(p)=b_x^\dagger(p)b_e(p) , \ \ \ \
\bar{J}_0(p)=\frac{1}{2}\left(b_e^\dagger(p)b_e(p)
-b_x^\dagger(p)b_x(p) \right).
\end{equation}
These operators also obey SU(2) commutation relations
\begin{equation}
[\bar{J}_+(p),\bar{J}_-(q)]=2\delta(p-q)\bar{J}_0(p) , \ \ \
[\bar{J}_0(p),\bar{J}_\pm(q)]=\pm\delta(p-q)\bar{J}_\pm(p)
\end{equation}
and they commute with the neutrino operators
\begin{equation}
[J_i(p),\bar{J}_j(q)]=0.
\end{equation}

Written in terms of $J_i(p)$ and $\bar{J}_i(p)$, the Hamiltonian
in Eq. (\ref{c1}) takes the following form:
\begin{eqnarray}\label{c4}
&&H_\nu+H_{\bar{\nu}}+H_{\nu\nu}+H_{\bar{\nu}\bar{\nu}}+H_{\nu\bar{\nu}}\\
&=&\int d^3p\left(\frac{\delta m^2}{2p}\cos{2\theta}-\sqrt{2}G_F
N_e\right) J_0(p) + \frac{1}{2}\int d^3p \frac{\delta
m^2}{2p}\sin{2\theta} \left(J_+(p)+J_-(p)\right) \nonumber
\\ &+& \int d^3p \left(-\frac{\delta m^2}{2p}\cos{2\theta}-\sqrt{2}G_F
N_e\right) \bar{J}_0(p) + \frac{1}{2}\int d^3p \frac{\delta
m^2}{2p}\sin{2\theta}
\left(\bar{J}_+(p)+\bar{J}_-(p)\right)\nonumber
\\ &+&\frac{\sqrt{2}G_F}{V}\int d^3pd^3q
R_{pq}\left({\bf J}(p)\cdot {\bf J}(q) + {\bf \bar{J}}(p)\cdot
{\bf \bar{J}}(q)\right)\nonumber
\\ &-&\frac{\sqrt{2}G_F}{V}\int d^3pd^3q
R_{pq}\left(-2J_0(p)\bar{J}_0(q)+J_+(p)\bar{J}_-(q)
+J_-(p)\bar{J}_+(q)\right)\nonumber
\end{eqnarray}
where a term proportional to identity is omitted.

As we did in the previous section, we introduce the coherent
states\footnote{In these formulas $\bar{z}$ is an independent
complex number, not the complex conjugate of $z$. Complex
conjugates will be denoted by a star such as $z^*$ and
$\bar{z}^*$.}
\begin{equation}\label{c5}
|z(t),\bar{z}(t)\rangle=N\bar{N}e^{\int_{\mathcal{P}} d^3p
z(p,t)J_+(p)}e^{\int_{\bar{\mathcal{P}}}
d^3\bar{p}\bar{z}(\bar{p},t)\bar{J}_-(\bar{p})}|\phi\rangle.
\end{equation}
Here $|\phi\rangle$ is analogous to the state defined in Eq.
(\ref{a6}), \emph{i.e.}, all permitted $\nu_e$ and $\bar{\nu}_e$ 
states are
occupied and all other neutrino flavor states are
empty:
\begin{equation}\label{c6}
|\phi\rangle=\prod_{p\in {\mathcal{P}}}a_e^\dagger(p)
\prod_{\bar{p}\in {\bar{\mathcal{P}}}}
b_e^\dagger(\bar{p})|0\rangle .
\end{equation}
In the above formulas, ${\mathcal{P}}$ and
${\bar{\mathcal{P}}}$ denote the set of all allowed momentum
modes for neutrinos and antineutrinos,
respectively. In what follows, we
will drop the symbols ${\mathcal{P}}$ and
${\bar{\mathcal{P}}}$ from the formulas by adopting the
convention that the non-overlined quantities such as $p,q$ will
take values in ${\mathcal{P}}$ whereas the overlined quantities
such as $\bar{p},\bar{q}$ take values in ${\bar{\mathcal{P}}}$.

The constants
\begin{equation}\label{c7}
N=\exp \left( - \frac{1}{2} \int d^3p \log (1+|z(p,t)|^2) \right),
\quad \bar{N}=\exp \left( - \frac{1}{2} \int d^3\bar{p} \log
(1+|\bar{z}(\bar{p},t)|^2) \right)
\end{equation}
in Eq. (\ref{c5}) normalize the coherent states:
\begin{equation}
\langle z,\bar{z}|z,\bar{z}\rangle =1 .
\end{equation}

A path integral representation of the evolution operator can be
given in terms of these coherent states as
\begin{equation}\label{c8}
 \langle z'(t_f),\bar{z}'(t_f)|\,U\,|z(t_i),\bar{z}(t_i)\rangle=\int
D[z,z^*,\bar{z},\bar{z}^*] e^{iS[z,z^*,\bar{z},\bar{z}^*]}
\end{equation}
where the action functional $S[z,z^*,\bar{z},\bar{z}^*]$
is\footnote{Here the expectation values are calculated using the states
in Eq. (\ref{c5}): $$\langle {\mathcal O} \rangle =\langle
z(t),\bar{z}(t)|{\mathcal O}|z(t),\bar{z}(t)\rangle
$$.}
\begin{equation}\label{c9}
S[z,z^*,\bar{z},\bar{z}^*]=\int_{t_i}^{t_f}dt \langle
i\frac{\partial}{\partial t}-(H_\nu+H_{\bar{\nu}}+H_{\nu\nu} +
H_{\bar{\nu}\bar{\nu}}+H_{\nu\bar{\nu}})\rangle -i\ln \langle
z'(t_f),\bar{z}'(t_f)|z(t_f),\bar{z}(t_f)\rangle.
\end{equation}
Once more we can find the stationary path by solving the
Euler-Lagrange equations for
\begin{equation}\label{c10}
L[z,z^*,\bar{z},\bar{z}^*]= \langle i\frac{\partial}{\partial
t}-(H_\nu+H_{\bar{\nu}}+H_{\nu\nu}+H_{\bar{\nu}\bar{\nu}} +
H_{\nu\bar{\nu}})\rangle.
\end{equation}
The linear terms in the Lagrangian can be calculated using 
\begin{equation}\label{E1}
\langle J_+(p) \rangle = \langle J_-(p) \rangle ^*=
\frac{z^*(p,t)}{1+|z(p,t)|^2}, \quad\quad \langle J_0(p) \rangle =
-\frac{1}{2}\ \frac{1-|z(p,t)|^2}{1+|z(p,t)|^2},
\end{equation}
\begin{equation}\label{E2}
\langle \bar{J}_+(\bar{p}) \rangle = \langle \bar{J}_-(\bar{p})
\rangle ^*= \frac{\bar{z}(\bar{p},t)}{1+|\bar{z}(\bar{p},t)|^2},
\quad\quad \langle \bar{J}_0(p) \rangle = \frac{1}{2}\
\frac{1-|\bar{z}({\bar{p}},t)|^2}{1+|\bar{z}({\bar{p}},t)|^2},
\end{equation}
which are valid for $p\in \mathcal{P}$ and $\bar{p}\in
\bar{\mathcal{P}}$. When $p\not\in \mathcal{P}$ or
$\bar{p}\not\in \bar{\mathcal{P}}$ the expectation values are
zero. To calculate the quadratic terms we use the identities
\begin{equation}\label{E3}
\langle J_a(p)J^\dagger_a(q) \rangle =\langle J_a(p)
\rangle\langle J^\dagger_a(q) \rangle , \quad \quad %
\langle \bar{J}_a(\bar{p})\bar{J}^\dagger_a(\bar{q}) \rangle
=\langle \bar{J}_a(\bar{p}) \rangle\langle
\bar{J}^\dagger_a(\bar{q}) \rangle ,
\quad \quad %
\langle J_a(p)\bar{J}^\dagger_a(\bar{q}) \rangle =\langle J_a(p)
\rangle\langle \bar{J}^\dagger_a(\bar{q}) \rangle.
\end{equation}
Here $a=0,\pm$ and we assumed $p\neq q$ and $\bar{p}\neq \bar{q}$.
Also note that
\begin{equation}\label{T}
\langle i\frac{d}{dt}\rangle =i \left( \int
d^3p\frac{\dot{z}(p)z^*(p)}{1+|z(p)|^2} +\int d^3\bar{p}
\frac{\dot{\bar{z}}(\bar{p})\bar{z}^*(\bar{p})}{1+|\bar{z}(\bar{p})|^2}
\right) .
\end{equation}
Using Eqs. (\ref{E1})-(\ref{T}) Lagrangian can be found as
follows:
\begin{eqnarray}
L=L_\nu+L_{\bar{\nu}}&-&\frac{1}{4}\int d^3p
\frac{B(p)(1-|z(p)|^2)+2B_{ex}(p)z(p)+2B_{xe}(p)z^*(p)}{1+|z(p)|^2}
\\ &+&\frac{1}{4}\int d^3\bar{p}
\frac{B(\bar{p})(1-|\bar{z}(\bar{p})|^2)
+2B_{xe}(\bar{p})\bar{z}(\bar{p})+2B_{ex}(\bar{p})\bar{z}^*(\bar{p})}
{1+|\bar{z}(\bar{p})|^2} . \nonumber
\end{eqnarray}
Here $L_\nu$ is the Lagrangian given in Eq. (\ref{b15a}) and
$L_{\bar{\nu}}$ is the same as $L_\nu$ except that we substitute
$\bar{z}(\bar{p})$ in place of $z(p)$ and change the sign of
$N_e$. In the above equation we also defined
\begin{equation}\label{B}
B(p)=\frac{\sqrt{2}G_F}{V} \int d^3q R_{pq}
\frac{1-|z(q,t)|^2}{1+|z(q,t)|^2} -\frac{\sqrt{2}G_F}{V} \int
d^3\bar{q}
R_{p\bar{q}}\frac{1-|\bar{z}(\bar{q},t)|^2}{1+|\bar{z}(\bar{q},t)|^2}
\end{equation}
and
\begin{equation}\label{B_ex}
B_{xe}(p)=B^*_{ex}(p)=2\frac{\sqrt{2}G_F}{V} \int d^3q
R_{pq}\frac{z(q,t)}{1+|z(q,t)|^2}- 2\frac{\sqrt{2}G_F}{V} \int
d^3\bar{q} R_{p\bar{q}}
\frac{\bar{z}^*(\bar{q},t)}{1+|\bar{z}(\bar{q},t)|^2} .
\end{equation}
Equations of motion which result from this Lagrangian are as
follows:
\begin{equation}\label{c12}
i\dot{z}(p,t)=\beta(p,t)-\alpha(p,t) z(p,t)-\beta^*(p,t) z^2(p,t)
,
\end{equation}
\begin{equation}\label{c13}
i\dot{\bar{z}}(\bar{p},t)=-\bar{\beta}^*(\bar{p},t)+\bar{\alpha}(\bar{p},t)
\bar{z}(\bar{p},t)+\bar{\beta}(\bar{p},t) \bar{z}^2(\bar{p},t)
\end{equation}
and the complex conjugates of these equations. The coefficients
$\alpha$ and $\beta$ are given by:
\begin{equation}
\alpha(p,t)=-\frac{\delta m^2}{2p}\cos{2\theta}+\sqrt{2} G_F
N_e+B(p) \quad \quad \beta(p,t)=\frac{1}{2} \frac{\delta
m^2}{2p}\sin{2\theta} + B_{ex}(p).
\end{equation}
The coefficients $\bar{\alpha}$ and $\bar{\beta}$ are the same
except that the sign of the $\delta m^2$ is changed.

We can again write the parameters $z$ and $\bar{z}$ in terms of
the one-body
neutrino and antineutrino wave functions $\psi_e$, $\psi_x$,
$\bar{\psi}_e$ and $\bar{\psi}_x$ as follows:
\begin{equation}\label{c20}
z(p,t)=\frac{\psi_x(p,t)}{\psi_e(p,t)} \ \ \ \ \ \ \mbox{and} \ \
\ \ \
\bar{z}(\bar{p},t)=\frac{\bar{\psi}^*_x(\bar{p},t)}{\bar{\psi}^*_e(\bar{p},t)}.
\end{equation}
We also have the normalization conditions
\begin{equation}\label{c21}
|\psi_e|^2+|\psi_x|^2=1 \ \ \ \ \ \ \mbox{and} \ \ \ \ \ \
|\bar{\psi}_e|^2+|\bar{\psi}_x|^2=1 .
\end{equation}
Substituting  Eq. (\ref{c20}) into Eqs. (\ref{c12}) and (\ref{c13})
we see that they are equivalent to the Schrodinger equations
\begin{equation}\label{c22}
i\frac{\partial}{\partial t} \left( \begin{array}{c} \psi_e(p,t)
\\ \psi_x(p,t) \\ \end{array} \right) =
\frac{1}{2}\left(%
\begin{array}{cc}
  A+B-\Delta\cos{2\theta} & B_{ex}+\Delta\sin{2\theta} \\
  B_{x e}+\Delta\sin{2\theta} & -A-B+\Delta\cos{2\theta} \\
\end{array} \right)
\left( \begin{array}{c} \psi_e(p,t) \\
\psi_x(p,t) \\ \end{array} \right),
\end{equation}
and
\begin{equation}\label{c23}
i\frac{\partial}{\partial t} \left( \begin{array}{c}
\bar{\psi}_e(\bar{p},t)
\\ \bar{\psi}_x(\bar{p},t) \\ \end{array} \right) =
\frac{1}{2}\left(%
\begin{array}{cc}
  A+B+\Delta\cos{2\theta} & B_{ex}-\Delta\sin{2\theta} \\
  B_{x e}-\Delta\sin{2\theta} & -A-B-\Delta\cos{2\theta} \\
\end{array} \right)
\left( \begin{array}{c} \bar{\psi}_e(\bar{p},t) \\
\bar{\psi}_x(\bar{p},t) \\ \end{array} \right).
\end{equation}
In the above formula $B$ and $B_{ex}$ are obtained by substituting
Eq. (\ref{c21}) into Eqs. (\ref{B}) and (\ref{B_ex}), {\emph i.e.},
\begin{equation}\label{c24}
B=\frac{\sqrt{2}G_F}{V} \int d^3q
R_{pq}\left(|\psi_e(q,t)|^2-|\psi_x(q,t)|^2\right)-\frac{\sqrt{2}G_F}{V}
\int d^3\bar{q} R_{p\bar{q}}
\left(|\bar{\psi}_e(\bar{q},t)|^2-|\bar{\psi}_x(\bar{q},t)|^2\right)
,
\end{equation}
\begin{equation}\label{c25}
B_{ex}=B^*_{xe}=2\frac{\sqrt{2}G_F}{V} \int d^3q
R_{pq}\psi_e(q,t)\psi^*_x(q,t) -2\frac{\sqrt{2}G_F}{V} \int
d^3\bar{q} R_{p\bar{q}}
\bar{\psi}_e(\bar{q},t)\bar{\psi}^*_x(\bar{q},t) .
\end{equation}

\section{Three Neutrino Flavors}

In this section we generalize our formalism to three neutrino
flavors, i.e. when $\theta_{13} \neq 0$. SU(3)
symmetry of the neutrinos can be represented by
the operators
\begin{equation}\label{Define T}
T_{ij}(p)=a_i^\dagger(p)a_j(p) \quad\quad \mbox{and} \quad\quad
\bar{T}_{ij}(p)=b_j^\dagger(p)b_i(p)
\end{equation}
where $i$ and $j$ run over the flavor indices $e,\mu$ and $\tau$.
These operators generate orthogonal SU(3) algebras:
\begin{eqnarray}\label{U(3) algebra}
[T_{ij}(p),T_{kl}(q)] &=&
\delta(p-q)\left(\delta_{kj}T_{il}(p)-\delta_{il}T_{kj}(p)\right)
,
\\ \nonumber [\bar{T}_{ij}(p),\bar{T}_{kl}(q)]
&=&-\delta(p-q)\left(\delta_{kj}\bar{T}_{il}(p) -
\delta_{il}\bar{T}_{kj}(p)\right) ,
\\ \nonumber [T_{ij}(p),\bar{T}_{kl}(q)] &=&0 .
\end{eqnarray}

The effective Hamiltonian describing the propagation of neutrinos
in a background of electrons is given by \cite{Sawyer:2005jk}
\begin{eqnarray}\label{Hamiltonian}
H&=& \sum_{i,j}\int d^3p \left[
\left(\gamma_{ij}(p)+\omega_{ij}(p)\right) T_{ij}(p)
+\left(\gamma_{ij}(p)-\omega_{ij}(p)\right) \bar{T}_{ij}(p)\right]\\
&+& \frac{G_F}{\sqrt{2}V}\int d^3 p d^3 q
R_{pq}\sum_{i,j}
(T_{ij}(p)-\bar{T}_{ij}(p))(T_{ji}(q)-\bar{T}_{ji}(q)).\nonumber
\end{eqnarray}
In this Hamiltonian, the coefficients $\gamma_{ij}(p)$ are the
elements of the symmetric, traceless matrix $\Gamma$ which is
given below:
\begin{equation}\label{Lambda}
\Gamma=
\left(\gamma_{ij}\right) %
=\frac{1}{3}Q_{23}Q_{13}Q_{12}%
\left( \begin{array}{ccc}
  \frac{-\delta m_{21}^2-\delta m_{31}^2}{2p} & 0 & 0 \\
  0 & \frac{\delta m_{21}^2-\delta m_{32}^2}{2p} & 0 \\
  0 & 0 & \frac{\delta m_{31}^2+\delta m_{32}^2}{2p} \\
\end{array}\right)%
Q^\dagger_{12}Q^\dagger_{13}Q^\dagger_{23} . %
\end{equation}
Here $Q_{12}$, $Q_{13}$ and $Q_{23}$ are neutrino mixing matrices
in vacuum:
\begin{equation}
\label{mixing}
Q_{23}Q_{13}Q_{12} = \left(\matrix{
     1 & 0 & 0  \cr
     0 &  C_{23}  & S_{23} \cr
     0 & - S_{23} &  C_{23} }\right)
 \left(\matrix{
     C_{13} & 0 &  S_{13}^{\ast} \cr
     0 &  1 & 0 \cr
     - S_{13} & 0&  C_{13} }\right)
 \left(\matrix{
     C_{12} & S_{12} &0 \cr
     - S_{12} & C_{12} & 0 \cr
     0 & 0&  1 }\right) ,
\end{equation}
where $C_{13}$, etc. is the short-hand notation for $\cos
{\theta_{13}}$, etc. Since $S_{13}$ may be multiplied by a phase,
we explicitly indicated its complex conjugate.
Note that individual matrices, not their matrix
elements, are called $Q_{23}$, $Q_{13}$, and $Q_{12}$, respectively in
Eq. (\ref{mixing}).

The terms in the Hamiltonian which are proportional to
$\gamma_{ij}$ represent vacuum oscillations of the neutrinos.
The coefficients $\omega_{ij}$ are real and symmetric. They
are the elements of the diagonal, traceless matrix $\Omega$ given
below:
\begin{equation}\label{Omega}
\Omega=\left(\omega_{ij}\right)=\frac{1}{3}
\left(%
\begin{array}{ccc}
  2V_c & 0 & 0 \\
  0 & -V_c & 0 \\
  0 & 0 & -V_c \\
\end{array}
\right) .
\end{equation}
Here $V_c$ is the Wolfenstein potential described earlier.

A path integral formula for the evolution operator of the system
can be constructed using SU(3) coherent states which are given by
\begin{equation}\label{Coherent states}
|z,\bar{z}\rangle = N\bar{N}%
e^{\int_{\mathcal{P}} dp\left(z_{\mu}(p)T_{\mu e}(p) +
z_{\tau}(p)T_{\tau e}(p)\right)}%
e^{\int_{\bar{\mathcal{P}}}d\bar{p}
\left(\bar{z}_{\mu}(\bar{p})\bar{T}_{e \mu}(\bar{p})
+\bar{z}_{\tau}(\bar{p})\bar{T}_{e \tau}(\bar{p})\right)}
|\phi\rangle .
\end{equation}
These coherent states are defined with respect to the reference
state $|\phi\rangle$ which is defined as in Eq. (\ref{c6}). The
normalization constants $N$ and $\bar{N}$ are given by
\begin{equation}
N=\exp\left( - \frac{1}{2} \int d^3p \log
(1+|z_{\mu}(p,t)|^2+|z_{\tau}(p,t)|^2) \right) ,
\quad \bar{N}=\exp \left(
- \frac{1}{2} \int d^3\bar{p} \log
(1+|\bar{z}_{\mu}(\bar{p},t)|^2+|\bar{z}_{\tau}(\bar{p},t)|^2) \right) .
\end{equation}

Neutrino SU(3) coherent states are characterized by two complex numbers
that we denoted by $z_{\mu}$ and $z_{\tau}$ in Eq.
(\ref{Coherent states}).
We used $\bar{z}_{\mu}$ and $\bar{z}_{\tau}$ for the SU(3) symmetry of the
antineutrinos. As a practical convention, we also define
$z_e(p)=\bar{z}_e(\bar{p})=1$ in what follows\footnote{For example
$$
\sum_{k}|z_k(p)|^2= 1+|z_{\mu}(p)|^2+|z_{\tau}(p)|^2.$$}.

Evolution operator can be given by the following path integral
formula in terms of the SU(3) coherent states:
\begin{equation}\label{Path Integral}
\langle z'(t_f),\bar{z}'(t_f)|U|z(t_i),\bar{z}(t_i)\rangle =%
\int D[z,\bar{z}] e^{iS[z,\bar{z}]},
\end{equation}
where the measure is given by Eq. (\ref{son}) of the Appendix.
In this formula
\begin{equation}\label{Lagrangian1}
S[z,\bar{z}]=\int_{t_i}^{t_f}dt\langle
z(t),\bar{z}(t)|i\frac{d}{dt}- H(t)|z(t),\bar{z}(t)\rangle
-i\ln\langle z'(t_f),\bar{z}'(t_f)|z(t_f),\bar{z}(t_f)\rangle
\end{equation}
plays the role of a classical action. The derivation of these
formulas and the exact expression for the integral measure can be
found in the Appendix.

As before, we write down the Lagrangian\footnote{Here we use the short
hand notation $\langle {\mathcal O} \rangle =\langle
z(t),\bar{z}(t)|{\mathcal O}|z(t),\bar{z}(t)\rangle $,
\emph{etc}.}
\begin{equation}
L[z,\bar{z}]=\langle i\frac{d}{dt}- H \rangle
\end{equation}
and solve the Euler-Lagrange equations
\begin{equation}\label{Euler Lagrange Equations}
\left(\frac{d}{dt}\frac{\partial}{\partial \dot{z}_n}-
\frac{\partial}{\partial z_n}\right) L[z,\bar{z}]=0  \ \ \
\mbox{and}  \ \ \ \left(\frac{d}{dt}\frac{\partial}{\partial
\dot{\bar{z}}_n}- \frac{\partial}{\partial \bar{z}_n}\right)
L[z,\bar{z}]=0
\end{equation}
for $n=\mu,\tau$ and those for $z_n^*$ and $\bar{z}_n^*$. We use
the expectation values of the SU(3) generators which are given
below:
\begin{equation}\label{Expectation values}
\langle T_{ij}(p)\rangle =
  \frac{z_i^*(p)z_j(p)}{\sum_{k}|z_k(p)|^2} ,
\quad \quad \langle %
\bar{T}_{ij}(\bar{p})\rangle =
\frac{\bar{z}_i(\bar{p})\bar{z}^*_j(\bar{p})}{\sum_{k}|\bar{z}_k(\bar{p})|^2}
\end{equation}
for $i,j=e,\mu,\tau$ (with the convention $z_e(p)=\bar{z}_e(p)=1$).
Here we assumed $p\in \mathcal{P}$ and $\bar{p}\in
\bar{\mathcal{P}}$. If $p\not\in \mathcal{P}$ or
$\bar{p}\not\in \bar{\mathcal{P}}$ then the expectation values
are zero. The quadratic terms in the Lagrangian are calculated
using the identities
\begin{equation}\label{Quadratic terms}
\langle T_{ij}(p)T_{ji}(q)\rangle =
\langle T_{ij}(p)\rangle \langle T_{ji}(q)\rangle ,  %
\quad\quad %
\langle \bar{T}_{ij}(\bar{p})\bar{T}_{ji}(\bar{q})\rangle
=\langle \bar{T}_{ij}(\bar{p})\rangle \langle \bar{T}_{ji}(\bar{q})\rangle , %
\end{equation}
and
\begin{equation}
\langle T_{ij}(p)\bar{T}_{ji}(\bar{q})\rangle =\langle
T_{ij}(p)\rangle \langle \bar{T}_{ji}(\bar{q})\rangle ,
\end{equation}
which are valid for $p\neq p$ and $\bar{p} \neq \bar{q}$. The
expectation value of the time derivative term is
\begin{equation}\label{Time derivative}
\langle i\frac{d}{dt}\rangle =i \sum_{n(\neq e)}%
\left( \int d^3 p\frac{\dot{z}_n(p)z^*_n(p)}{\sum_{k}|z_k(p)|^2}
+\int d^3 \bar{p} \frac{\dot{\bar{z}}_n
(\bar{p})\bar{z}^*_n(\bar{p})}{\sum_{k}|\bar{z}_k(\bar{p})|^2}
\right) .
\end{equation}
Using formulas (\ref{Expectation values})-(\ref{Time derivative})
we find the Lagrangian as
\begin{eqnarray}\label{Lagrangian2}
L[z,\bar{z}]&=&i \sum_{n(\neq e)} \left( %
\int d^3 p\frac{\dot{z}_n(p)z^*_n(p)}{\sum_{k}|z_k(p)|^2}+%
\int d^3 \bar{p}
\frac{\dot{\bar{z}}_n(\bar{p})\bar{z}^*_n(\bar{p})}{\sum_{k}|\bar{z}_k(\bar{p})|^2}\right)\\
&-& \sum_{i,j}\int d^3 p
\left(\gamma_{ij}(p)+\omega_{ij}(p)+\frac{1}{2}Y_{ij}(p)\right)
\frac{z_i^*(p)z_j(p)}{\sum_{k}|z_k(p)|^2} \nonumber \\
&-& \sum_{i,j}  \int d^3 \bar{p}
\left(\gamma_{ij}(\bar{p})-\omega_{ij}(\bar{p}) -
\frac{1}{2}Y_{ij}(\bar{p})\right)
\frac{\bar{z}_i(\bar{p})\bar{z}^*_j(\bar{p})}{\sum_{k}|\bar{z}_k(\bar{p})|^2}
. \nonumber
\end{eqnarray}
Here $Y_{ij}(p)$ is given by
\begin{equation}\label{Y_ij}
Y_{ij}(p)=\frac{\sqrt{2}G_F}{V}\left(%
\int d^3 q R_{pq} \frac{z_i(q)z_j^*(q)}{\sum_{k}|z_k(q)|^2} -\int
d^3 \bar{q} R_{p\bar{q}}
\frac{\bar{z}^*_i(\bar{q})\bar{z}_j(\bar{q})}{\sum_{k}|\bar{z}_k(\bar{q})|^2}
\right) .
\end{equation}
The equations of motion driven from this Lagrangian are given
below:
\begin{eqnarray}
i\dot{z}_n(p)&=&\sum_{i}\left[
\left(\gamma_{ni}(p)+\omega_{ni}(p)+Y_{ni}(p)\right)z_i(p)  %
-\left(\gamma_{ei}(p)+\omega_{ei}(p)+Y_{ei}(p)\right)z_n(p)z_i(p)   %
\right] , \label{Equation of motion 1}\\
i\dot{\bar{z}}_n(\bar{p})&=&\sum_{i}\left[
\left(\gamma^*_{ni}(\bar{p})-\omega_{ni}(\bar{p}) -
Y^*_{ni}(\bar{p})\right)\bar{z}_i(\bar{p})  %
-\left(\gamma^*_{ei}(\bar{p})-\omega_{ei}(\bar{p}) -
Y^*_{ei}(\bar{p})\right)\bar{z}_n(\bar{p})\bar{z}_i(\bar{p})   %
\right] , \label{Equation of motion 2}
\end{eqnarray}
where $n=\mu,\tau$. We write the parameters $z_i$ and $\bar{z}_i$
in terms of the neutrino wavefunctions as
$z_i(p)=\psi_i(p)/\psi_e(p)$ and
$\bar{z}_i(\bar{p})=\bar{\psi}^*_i(\bar{p})/\bar{\psi}^*_e(\bar{p})$,
\emph{i.e.},
\begin{equation}\label{Reparametrization 1}
z_e(p)=\frac{\psi_e(p)}{\psi_e(p)}=1, \quad
z_\mu(p)=\frac{\psi_\mu(p)}{\psi_e(p)}, \quad
z_\tau(p)=\frac{\psi_\tau(p)}{\psi_e(p)}
\end{equation}
and
\begin{equation}\label{Reparametrization 2}
\bar{z}_e(\bar{p})=\frac{\bar{\psi}^*_e(\bar{p})}{\bar{\psi}^*_e(\bar{p})}=1,
\quad
\bar{z}_\mu(\bar{p})=\frac{\bar{\psi}^*_\mu(\bar{p})}{\bar{\psi}^*_e(\bar{p})},
\quad \bar{z}_\tau(\bar{p}) =
\frac{\bar{\psi}^*_\tau(\bar{p})}{\bar{\psi}^*_e(\bar{p})} ,
\end{equation}
where the one body wavefunctions are normalized as follows:
\begin{equation}\label{normalization}
|\psi_e(p)|^2+|\psi_\mu(p)|^2+|\psi_\tau(p)|^2=1 , \quad\quad
|\bar{\psi}_e(\bar{p})|^2 +
|\bar{\psi}_\mu(\bar{p})|^2+|\bar{\psi}_\tau(\bar{p})|^2=1.
\end{equation}
If the Eqs. (\ref{Reparametrization 1}) and
(\ref{Reparametrization 2}) are substituted in Eqs. (\ref{Equation
of motion 1}) and (\ref{Equation of motion 2}), we find that
$\psi_i(p)$ and $\bar{\psi}_i(\bar{p})$ obey the following
Schrodinger equations:
\begin{equation}
i\frac{d}{dt} %
\left(\begin{array}{c}
  \psi_e(p) \\ \psi_\mu(p) \\ \psi_\tau(p) \\
\end{array} \right)
= \left( \Gamma(p)+\Omega(p) + Y(p) \right) %
\left( \begin{array}{c}
  \psi_e(p) \\ \psi_\mu(p) \\ \psi_\tau(p) \\
\end{array} \right) ,
\end{equation}
\begin{equation}
i\frac{d}{dt} %
\left(\begin{array}{c}
  \bar{\psi}_e(\bar{p}) \\ \bar{\psi}_\mu(\bar{p})
\\ \bar{\psi}_\tau(\bar{p}) \\
\end{array} \right)
= \left(-\Gamma(\bar{p})+\Omega(\bar{p}) + Y(\bar{p}) \right) %
\left( \begin{array}{c}
  \bar{\psi}_e(\bar{p}) \\ \bar{\psi}_\mu(\bar{p}) \\ \bar{\psi}_\tau(\bar{p})
\end{array} \right) .
\end{equation}
Here, the matrix $Y$ is the matrix formed by the elements $Y_{ij}$
given in Eq.(\ref{Y_ij}).
Note that as one goes from the neutrino to antineutrino equations
only the signs of the $\delta m^2$ terms change.
If we substitute Eq.
(\ref{Reparametrization 1}) and (\ref{Reparametrization 2}) in
Eq. (\ref{Y_ij}) we see that $Y_{ij}$ can be written in terms of
the one-body wavefunctions as
\begin{equation}
Y_{ij}(p) = \frac{\sqrt{2}G_F}{V}\left( \int d^3 q
R_{pq} %
 \psi_i(q)\psi^*_j(q) - \int d^3 \bar{q}
R_{p\bar{q}}  \bar{\psi_i}(\bar{q}) \bar{\psi}_j^*(\bar{q})\right)
.
\end{equation}

\section{Conclusions}

In this article an algebraic approach to the neutrino propagation in
dense media is presented. The Hamiltonian describing a gas of neutrinos
interacting with each other and background fermions is written in terms
of the appropriate SU(2) (for two flavors) or SU(3) (for three flavors)
operators. Neutrinos as well as antineutrinos are considered.
The evolution of the resulting many-body problem is formulated as 
either an SU(2) or an SU(3) coherent-state path integral. The evolution
operator for the entire system is calculated using two different
approximations, namely the saddle-point approximation and the operator
product linearization approximation. In our case these two approximations
yield the same answer. This approximate solution of the neutrino evolution
is the only one used so far in applications to the core-collapse supernovae
and the Early Universe. It is important to stress that this solution is only
an approximation to the many-body problem described by Eq. (\ref{c1}) 
and the corrections to it
may play a significant role.

We should also emphasize that the results for the evolution operator
obtained here are applicable to any initial state. An initial state such
as the one depicted in Eq. (\ref{a6}) represents an electron neutrino gas
described by a pure state, i.e. $\rho^2 = \rho$. 
Such an initial state may be a good description of
neutrinos produced in earlier stages of stellar evolution, but is 
inadequate for neutrinos in a thermal distribution. In such cases
it would be more appropriate to describe the neutrino gas using density
operators satisfying the equation
\begin{equation}\label{e2}
i\dot{\rho}=[H,\rho],
\end{equation}
where $H$ is the Hamiltonian of Eq. (\ref{c1}). The initial density matrix
$\rho_i$, then evolves as $\rho = U \rho_i U^{\dagger}$. Sometimes a
polarization vector is introduced. In our notation the one-body
polarization vector is defined as
\begin{equation}
\label{z1}
 P_i(q) = {\rm Tr} ( J_i(q) \rho)
\end{equation}
for two flavors. For three neutrino flavors one simply calculates the trace
above with the SU(3) generators, resulting in an eight-dimensional vector.
The polarization vector of Eq. (\ref{z1}) satisfies the equation
\begin{equation}
\label{z2}
i\dot{{\bf P}}(q) = {\rm Tr} ( [{\bf J}(q),H] \rho).
\end{equation}
It is straightforward to show that linearizing the Hamiltonian of
Eq. (\ref{c1}) using Eq. (\ref{rpa}) and substituting this
linearized Hamiltonian in Eq. (\ref{z2}) yields the evolution
equations of the polarization vectors as stated, e.g., in Refs.
\cite{Pastor:2002we} and \cite{Balantekin:2004ug}.

\section*{ACKNOWLEDGMENTS}
We thank G. Fuller, A. Malkus, R. Sawyer, 
and H. Yuksel for useful discussions.
This work was supported in part by the U.S. National Science
Foundation Grant No. PHY-0555231 and in part by the University of
Wisconsin Research Committee with funds granted by the Wisconsin
Alumni Research Foundation.

\section*{Appendix: Path Integral Representation of the Evolution
Operator}

In this appendix, we drive the SU(3) path integral formula in Eq.
(\ref{Path Integral}). The path integral formula for SU(2)
coherent states follows very similar lines.

The coherent states defined in Eq. (\ref{Coherent states}) admit
the following resolution of identity:
\begin{equation}\label{Resolution of identity}
I=\int %
\prod_p \frac{3!}{(2\pi i)^2}
\frac{d^2z_{\mu}(p)d^2z_{\tau}(p)}{\left(\sum_{k}
|z_k(p)|^2\right)^3}\prod_{\bar{p}} \frac{3!}{(2\pi i)^2}
\frac{d^2\bar{z}_{\mu}(\bar{p})d^2\bar{z}_{\tau}(\bar{p})}{\left(\sum_{k}
|\bar{z}_k(\bar{p})|^2\right)^3} |z,\bar{z}\rangle\langle
z,\bar{z}| .
\end{equation}

One can use this resolution of identity to write down the path
integral formula for the matrix element of the evolution operator
$U$. The procedure is well established \cite{koonin}
and will only be outlined 
in what follows: We start by dividing the time interval $[0,T]$
into $N$ infinitesimally small pieces:
\begin{equation}
t_0=0 ,\quad t_1=\varepsilon ,\quad t_2=2\varepsilon , \; \dots ,
\quad T=t_N=N\varepsilon .
\end{equation}
Ignoring the time dependence of the Hamiltonian in each
infinitesimal time interval we can write
\begin{equation}
\label{Matrix Element of U} \langle
z'(T),\bar{z}'(T)|U(T)|z(0),\bar{z}(0)\rangle =\langle
z'(T),\bar{z}'(T)|e^{-i\varepsilon H(t_{N})} e^{-i\varepsilon
H(t_{N-1})} \dots e^{-i\varepsilon H(t_1)} |z(0),\bar{z}(0)\rangle
.
\end{equation}
We then insert a resolution of identity to the left of each
exponential. We will put an additional label to $t_\alpha$ to the
variables $z_i(p)$ and $\bar{z}_i(p)$ in the resolution of
identity which is inserted just next to $e^{-i\varepsilon
H(t_\alpha)}$. This way, we obtain
\begin{eqnarray}\label{Matrix Element of U 2}
&&\langle z'(T),\bar{z}'(T)|U(T)|z(0),\bar{z}(0)\rangle =%
\int %
\prod_{\alpha=1}^N \prod_p\frac{3!}{(2\pi i)^2}
\frac{d^2z_{\mu}(p,t_\alpha)d^2z_{\tau}(p,t_\alpha)}{\left(\sum_{k}
|z_k(p,t_\alpha)|^2\right)^3} \prod_{\alpha=1}^N
\prod_{\bar{p}}\frac{3!}{(2\pi i)^2}
\frac{d^2\bar{z}_{\mu}(\bar{p},t_\alpha)d^2\bar{z}_{\tau}(\bar{p},t_\alpha)}
{\left(\sum_{k} |\bar{z}_k(\bar{p},t_\alpha)|^2\right)^3}\nonumber\\
&&\times\langle z'(T),\bar{z}'(T)|z(t_N),\bar{z}(t_N)\rangle
\prod_{\alpha=1}^N \langle
z(t_\alpha),\bar{z}(t_\alpha)|e^{-i\varepsilon
H(t_\alpha)}|z(t_{\alpha-1}),\bar{z}(t_{\alpha-1})\rangle .
\end{eqnarray}
Assuming that only the continuous paths will contribute to the
integral when we take $N\to\infty$ limit, we can write
\begin{equation}
|z(t_{\alpha-1}),\bar{z}(t_{\alpha-1})\rangle
=\left(1-\varepsilon\frac{d}{dt}\right)|z(t_{\alpha}),
\bar{z}(t_{\alpha})\rangle .
\end{equation}
In this case the infinitesimal propagator $\langle
z(t_\alpha),\bar{z}(t_\alpha)|e^{-i\varepsilon
H(t_\alpha)}|z(t_{\alpha-1}),\bar{z}(t_{\alpha-1})\rangle $ can be
written as
\begin{equation}
\langle z(t_\alpha),\bar{z}(t_\alpha)|(1-i\varepsilon
H(t_\alpha))(1-\varepsilon\frac{d}{dt})|z(t_{\alpha}),
\bar{z}(t_{\alpha})\rangle =e^{i\varepsilon\langle
z(t_\alpha),\bar{z}(t_\alpha)|(i\frac{d}{dt}-
H(t_\alpha))|z(t_{\alpha}),\bar{z}(t_{\alpha})\rangle } .
\end{equation}
Substituting this into Eq.(\ref{Matrix Element of U 2}) and taking
the limits $N\to\infty$ and $\varepsilon\to 0$ such that
$N\varepsilon=T$ we arrive at the following path integral formula
for the propagator:
\begin{equation}
\langle z'(T),\bar{z}'(T)|U(T)|z(0),\bar{z}(0)\rangle =%
\int D[z,\bar{z}] e^{i S[z,\bar{z}]} .
\end{equation}
In this formula $S[z,\bar{z}]$ is given by
\begin{equation}
S[z,\bar{z}]=\int_0^T dt \langle z(t),\bar{z}(t)|i\frac{d}{dt}-
H(t)|z(t),\bar{z}(t)\rangle -i\log \langle
z'(T),\bar{Z}'(T)|z(T),\bar{z}(T)\rangle
\end{equation}
and the path integral measure is
\begin{eqnarray}
\label{son}
D[z,\bar{z}]&=&\lim_{N\to\infty}%
e^{-3\sum_{\alpha=1}^N\int dp %
\log \left(\sum_{k}|z_k(p,t_\alpha)|^2\right)}
e^{-3\sum_{\alpha=1}^N\int d\bar{p} %
\log \left(\sum_{k}|\bar{z}_k(\bar{p},t_\alpha)|^2\right)} \nonumber\\
&&\prod_{\alpha=1}^N \prod_p\frac{3!}{(2\pi i)^2}
d^2z_{\mu}(p,t_\alpha)d^2z_{\tau}(p,t_\alpha)\prod_{\alpha=1}^N
\prod_{\bar{p}}\frac{3!}{(2\pi i)^2}
d^2\bar{z}_{\mu}(\bar{p},t_\alpha)d^2\bar{z}_{\tau}(\bar{p},t_\alpha)
.
\end{eqnarray}


\end{document}